\documentclass[aps,prb,twocolumn,showpacs,superscriptaddress,floatfix,reprint]{revtex4-2}

\usepackage{amsmath,amssymb}
\usepackage{graphicx}
\usepackage{dcolumn}
\usepackage{bm}
\usepackage{hyperref}

\usepackage{comment}
\usepackage{xcolor}

\begin{document}

\preprint{APS/123-QED}

\title{Self-induced Transparency in a Semiconductor Quantum Dot medium at ultra-cold temperatures}

\author{Samit Kumar Hazra}
\email{samit176121009@iitg.ac.in}
\affiliation{Department of Physics, Indian Institute of Technology Guwahati, Guwahati 781039, Assam, India}
\author{P. K. Pathak}
\email{ppathak@iitmandi.ac.in}
\affiliation{Indian Institute of Technology Mandi, Mandi 175001, Himachal Pradesh, India}
\author{Tarak Nath Dey}
\email{tarak.dey@iitg.ac.in}
\affiliation{Department of Physics, Indian Institute of Technology Guwahati, Guwahati 781039, Assam, India}

\date{\today}
\begin{abstract}
We investigate the feasibility of minimum absorption and minimum broadening of pulse propagation in an inhomogeneously broadened semiconductor quantum dot medium. The phonon interaction is inevitable in studying any semiconductor quantum dot system. We have used the polaron transformation technique to deal with quantum dot phonon interaction in solving system dynamics. We demonstrate that a short pulse can propagate inside the medium with minimal absorption and broadening in pulse shape. The stable pulse area becomes slightly higher than the prediction of the pulse area theorem and is also dependent on the environment temperature. The change in the final pulse shape is explained very well by numerically solving the propagation equation supported by the susceptibility of the medium. Our system also exhibits the pulse breakup phenomena for higher input pulse areas. Therefore, the considered scheme can have important applications in quantum communication, quantum information, and mode-locking with the advantage of scalability and controllability.
\end{abstract}

\maketitle


\section{INTRODUCTION}
\label{sec:intro}
In Self-induced transparency (SIT), an optical pulse propagates resonantly through the two-level absorbing medium without any loss and distortion. This pioneering work was carried out by McCall and Hahn \cite{McHa1, McHa2}. SIT originates from the generated coherence of a strongly coupled light-medium interaction. Therefore for observing SIT, the incident pulse should be short compared to the various relaxation times present in the system, such that the coherence will not vanish during the pulse propagation. Further, the pulse should also be strong enough to excite the atom from the ground state.  
 One of the best theoretical estimations of the input pulse was reported in the ``area theorem" \cite{McHa2}. This theorem dictates that a 2$\pi$ secant pulse can propagate through the medium without any loss and distortion in the pulse shape. In general, for an initial pulse area $\theta_{0}$ obeying the condition $(n+1)\pi >\theta_{0}>n\pi$, evolves the area towards $(n+1)\pi$ or $n\pi$ depending on whether $n$ is odd or even. Therefore input pulse with a larger area of $2n\pi$, breaks up into $n$ number of $2\pi$ pulses with different propagation velocities. These effects have been observed experimentally in atomic rubidium medium by Slusher and Gibbs \cite{SluGib1}. In particular, they have found excellent agreement between numerical simulations and experimental results. These fundamental properties of the SIT were investigated several times, both theoretically and experimentally \cite{Allen1, Lamb1, Eilbeck1}.

However, in atomic medium, the preparation and trapping of atomic gas required a vast and sophisticated setup. Moreover, due to the gaseous nature of the medium, the different velocity of the atom shows Dopler broadening in output result.
For the last two decades, solid-state semiconductor mediums have emerged as a potential candidate for optical applications, particularly for scalable on-chip quantum technology. Earlier, the resonant coherent pulse propagation in bulk and quantum-well semiconductors behaves differently compared to a two-level atomic medium. The discrepancy mentioned above occurs due to the many-body Coulomb interaction of the different momentum states present in a bulk medium\cite{SWKoch1, HG1, ASch1}. This problem has been overcome in three-dimensionally confined excitons in quantum dots (QD's). The quantum dots can easily be engineered to get the desired transition frequency to avoid the problem of laser availability. The scalability and fabrication technology make the semiconductor QDs suitable for modern quantum optics experiments. There have been some interesting theoretical proposals about the possibility of observing SIT in self-organized InGaAs QDs\cite{Gpan1}. Excitonic transition in InGaAs QDs have large transition dipole moments and long dephasing time in the range of nanoseconds at cryogenic temperatures \cite{PBor1} and are, therefore a promising candidate for SIT.

Though the QD medium is a potential candidate for observing SIT, it has a few drawbacks also. All the QDs inside the medium are not identical, so an inhomogeneous level broadening is always present in the system. In semiconductors,  longitudinal acoustic phonon interaction is vital because of the environment temperature. Interactions between phonon and exciton lead to dephasing in coupled dynamics of exciton-photon interaction\cite{DPHZ1,DPHZ2}. Several theoretical models and experiments have recently explained SIT in the semiconductor QD medium \cite{QDSIT1, QDSIT2, QDSIT3}. Few of them consider the effect of the phonon environment on the system dynamics in the context of group velocity dispersion\cite{Phonon1}. Another recent experimental work showed the SIT mode-locking and area theorem for semiconductor QD medium, and rubidium atom \cite{MODELOCK1, MODELOCK2}.

In this paper, we discuss the possibility of SIT in a semiconductor QD medium incorporating the effect of phonon bath in our model. We utilize the recently developed polaron transformed master equation keeping all orders of exciton-phonon interaction \cite{POLARON1, POLARON2, POLARON3}. Our model's pulse propagation dynamics depend on system and bath parameters. Hence, the propagation dynamics become more transparent by knowing both the system and the bath's contribution.The motivation behind this work is to find long-distance optical communication without loss of generality in an array of QD. Due to strong confinement of electron hole pairs, QDs have discrete energy levels thus QD arrays mimic atomic medium with the added advantage of scalability and controllability with advanced semiconductor technology. It is also possible to create QD fibers which can be used for quantum communication channels \cite{QDFIBER,QCOM}. Motivated by this work, we theoretically investigate the self-induced transparency effect in a semiconductor QD medium.

Our paper is organized as follows. Sec.\ \ref{sec:intro} contains a brief introduction of the SIT in a QD medium and its application.
In Sec.\ \ref{sec:model}, we present our considered model system along with the theoretical formalism of the polaron master equation. In Sec.\ \ref{sec:result} we discuss the result after numerically solving the relevant system equations. Finally, we draw a conclusion in Sec.\ \ref{sec:conclud}.

\section{MODEL SYSTEM}
\label{sec:model}

\begin{figure}[t]
   \includegraphics[scale=0.45]{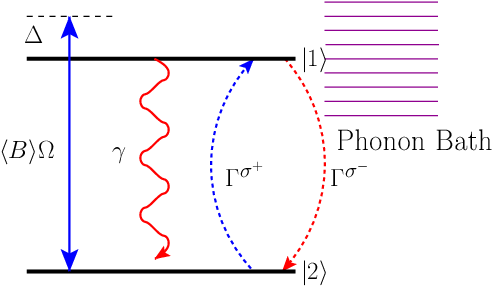}
    \caption{A Schematic diagram of the QD level system with ground state $\vert 2 \rangle$ and exciton state $\vert 1 \rangle$ driven by the optical pulse with effective coupling $\langle B\rangle\Omega$(blue line). The spontaneous decay from the exciton state to the ground state is shown using a curly red line. The parallel violet lines represent the phonon modes interacting with the exciton state. The red and blue dashed lines represent the phonon-induced decay and pumping rate respectively.}
    \label{Fig.1}
\end{figure}

The phonon contribution to QD dynamics at low temperature is mandatory. We assume the propagation of an optical pulse along the $z$-direction. Accordingly, we define the electric field of the incident optical pulse as
\begin{equation}
\vec{E}(z,t) = \hat{e}\mathcal{E}(z,t)e^{i(kz-\omega_{L} t)} + c.c ,
\end{equation}
where $\mathcal{E}(z,t)$ is the slowly varying envelope of the field. The bulk QD medium comprises multiple alternating InGaAs/GaAs QD deposition layers. Every QD inside the medium strongly interacts with the electric field due to the significant dipole moment.
Since all the QD inside the medium is not identical, the exciton energy of the different  QD will vary depending on the dot size. The $l^{th}$ type QD can be modeled as a two-level system with exciton state $\vert 1\rangle _{l}$, and ground state $\vert 2\rangle _{l}$ with energy gap $\hbar\omega_{l}$ by taking the proper choice of biexciton binding energy and polarisation as shown in the Fig.\ref{Fig.1}. The raising and lowering operator for the $l^{th}$ type QD can be written as $\sigma^{+}_{l} = \vert 1\left(\omega_{l}\right)\rangle _{l}\langle 2\left(\omega_{l}\right)\vert _{l}$ and $\sigma^{-}_{l} = \vert 2\left(\omega_{l}\right)\rangle _{l}\langle 1\left(\omega_{l}\right)\vert _{l}$.

 In case of semiconductor QD's, the optical properties get modified due to the lattice mode of vibration \textit{i.e.}, the acoustic phonon. Hence, QD exciton transition coupled to an acoustic phonon bath model mimics the desired interaction. The phonon bath consists of a large number of closely spaced harmonic oscillator modes. Therefore, we introduce the annihilation and creation operators associated with $k^{th}$ phonon mode having frequency $\omega_{k}$ as $b_{k}$ and $b_{k}^{\dagger}$. The mode frequency can be expressed as $\omega_{k} = c_{s}k$ where $k$ and $c_{s}$ are the wave vector and velocity of sound. The Hamiltonian for the described model system after making dipole and rotating wave approximation is given by
\begin{equation}
\begin{split}
H &= \sum\limits_{l}\Bigr[- \hbar\delta_{l}\sigma^{+}_{l}\sigma^{-}_{l} + \frac{1}{2}\hbar\Bigl( \Omega(z,t)\sigma^{+}_{l} + \Omega^{*}(z,t)\sigma^{-}_{l} \Bigr) \\
  &+ \hbar\sigma^{+}_{l}\sigma^{-}_{l}\sum\limits_{k}\lambda_{k}\left(b_k+b_k^{\dag}\right)   
  \Bigr] +\hbar\sum\limits_k\omega_kb_k^{\dag}b_k,
\end{split}
\end{equation}
where $\lambda_{k}$ is the exciton phonon mode coupling constant and $\Omega(z,t) = - 2 \vec{d}_{12}\cdot\hat{e}\mathcal{E}(z,t)/\hbar$ is the Rabi frequency with transition dipole moment vector $\vec{d}_{12}$. The detuning of the optical field with QD transition is defined as $\delta_{l} = \omega_{L} - \omega_{l}$.

We notice that the Hamiltonian contains an infinite sum over phonon modes. Keeping all order of exciton phonon interaction, we made a transformation in the polaron frame. The transformation rule for modified Hamiltonian is given by  $H^{'} = e^{P} H e^{-P}$ where the operator $P = \sum_{l}\sigma^{+}_{l}\sigma^{-}_{l}\sum_{k}\lambda_{k}( b_{k}^{\dagger} - b_{k})/\omega_{k}$. This transformation also helps us to separate the system Hamiltonian from the total Hamiltonian which is our primary interest. The transformed Hamiltonian is divided into system, bath and interaction part, which can be decomposed as $H^{\prime}=H_s+H_b+H_{I}$, where
\begin{eqnarray}
H_{s} &=& \sum\limits _{l} - \hbar\Delta _{l}\sigma^{+}_{l}\sigma^{-}_{l} + \langle B\rangle X_{l}^{g},\\
H_b &=& \hbar\sum_k\omega_k b_k^{\dag}b_k,\\
H_{I} &=& \sum\limits _{l}\xi_g X_l ^{g}+\xi_u X_l ^{u},
\end{eqnarray}
 and $\Delta_{l}$ is the redefined detuning by considering the polaron shift $\sum_k\lambda_{k}^2/\omega_k$.
The definition of phonon-modified system operators is given by 
\begin{eqnarray}
X_{l}^{g} &=& \frac{\hbar}{2}\left(\Omega(z,t)\sigma^{+}_{l} + \Omega^{*}(z,t)\sigma^{-}_{l} \right),\\
X_{l}^{u} &=& \frac{i\hbar}{2}\left(\Omega(z,t)\sigma^{+}_{l} - \Omega^{*}(z,t)\sigma^{-}_{l} \right).
\end{eqnarray}
The phonon bath fluctuation operators are
\begin{eqnarray}
 \xi_{g} &=& \frac{1}{2}\left( B_{+} + B_{-} -2\langle B\rangle  \right),\\
 \xi_{u} &=& \frac{1}{2i}\left( B_{+} - B_{-} \right),
 \end{eqnarray}
where $B_{+}$ and $B_{-}$ are the coherent-state phonon  displacement operators. Explicitly, the phonon displacement operators in terms of the phonon mode operators can be written as
\begin{equation}
B_{\pm} =  \exp\left[{\pm \sum_{k} \frac{\lambda_{k}}{\omega_{k}}\left( b_{k}^{\dagger} - b_{k}\right)}\right].\nonumber 
\end{equation}
From this expression, it is clear that the exponential of the phonon operator takes care of all the higher order phonon processes.
Therefore, the phonon displacement operator averaged over all closely spaced phonon modes at a temperature T, obeys the relation $\langle B_{+}\rangle =\langle B_{-}\rangle = \langle B\rangle$ where
\begin{equation}
\langle B\rangle= \text{exp}\left[-\frac{1}{2}\int_0^{\infty}d\omega\frac{J(\omega)}{\omega^2}
\coth\left(\frac{\hbar\omega}{2K_{B}T}\right)\right],
\end{equation}
and $K_{B}$ is the Boltzmann constant.
The phonon spectral density function $J(\omega)= \alpha_{p}\omega^3\exp[-\omega^2/2\omega_b^2]$  describes longitudinal acoustic(LA) phonon coupling via a deformation potential \cite{SPECT} for QD system, where the parameters $\alpha_p$ and $\omega_b$ are the electron-phonon coupling and cutoff frequency, respectively.

Next we use the master equation(ME) approach to solve the polaron-transformed system Hamiltonian dynamics by considering the phonon bath as a perturbation. The Born-Markov approximation can be performed with respect to the polaron-transformed perturbation in the case of nonlinear excitation. Hence, the density matrix equation for the reduced system under Born-Markov approximation can be written as
\begin{equation}
\dot{\rho} = \frac{1}{i\hbar}[H_s,\rho]+\sum\limits_{l}\Bigr({\cal L}_{ph}\rho
+\frac{\gamma}{2}{\cal L}[\sigma^{-}_{l}]\rho
 +\frac{\gamma_d}{2}{\cal L}[\sigma^{+}_{l}\sigma^{-}_{l}]\rho \Bigr)\label{meq},
\end{equation}
where $\gamma$ is the spontaneous decay rate of the exciton state. The spontaneous decay originates from the quantum fluctuations of the vacuum state. Similarly, for thermal fluctuation, we have adopted the final Lindbladian form of the dephasing interaction model described by a simple stochastic Hamiltonian\cite{GSA}. Therefore, we incorporate the pure-dephasing process phenomenologically in ME with a decay rate $\gamma_d$. This additional dephasing term explains the broadening of the zero-phonon line (ZPL) in QD with increasing temperatures \cite{ZPL1,ZPL2}. The Lindblad superoperator $\cal L$ is expressed as  ${\cal L}[\cal O]\rho = \text{2} {\cal O} \rho {\cal} O^{\dagger} - {\cal O}^{\dagger}{\cal O} \rho - \rho \cal O^{\dagger}\cal O $, under the operation of $\cal O$ operator. The term ${\cal L}_{ph}$ represents the effect of phonon bath on the system dynamics. Therefore the explicit form of ${\cal L}_{ph}\rho$ in terms of previously defined  system operators can be expressed as
\begin{eqnarray}
{\cal L}_{ph}\rho &=& -\frac{1}{\hbar^2}\int_0^{\infty}d\tau\sum_{j=g,u}G_j(\tau)[X_l^{j}(z,t),X_l^{j}(z,t,\tau)\rho(t)]~\nonumber\\
 &+& H.c.,
\end{eqnarray}
where $X_l^{j}(z,t,\tau)=e^{-iH_s\tau/\hbar}X_l^{j}(z,t)e^{iH_s\tau/\hbar}$, and the polaron Green's functions are $G_g(\tau)=\langle B\rangle^2\{\cosh\left[\phi(\tau)\right]-1\}$ and $G_u(\tau)=\langle B\rangle^2\sinh[\phi(\tau)]$. 
The phonon Green's functions depend on phonon correlation function given below 
\begin{equation}
\phi(\tau)=\int_0^{\infty}d\omega\frac{J(\omega)}{\omega^2}
\left[\coth\left(\frac{\hbar\omega}{2K_{B}T}\right)\cos(\omega\tau)-i\sin(\omega\tau)\right].\label{phi}
\end{equation}
The polaron ME formalism is not generally valid for arbitrary excitation strength and exciton phonon coupling. The validity of polaron ME is stated as \cite{POLARON1}
\begin{equation}
      \left( \frac{\Omega}{\omega_{b}} \right)^{2} \left(1 - \langle B\rangle^{4}\right) \ll 1 .\label{valid}
\end{equation}
It is clear from the above equation that, at low temperatures $\langle B\rangle \approx 1$ and $\Omega/\omega_{b} < 1$ fulfil the above criteria. Hence, we restrict our calculation in the weak field regime satisfying $\Omega/\omega_{b} < 1$   at a low phonon bath temperature.\\

The full polaron ME (\ref{meq}) contains multiple commutator brackets and complex operator exponents, which require involved numerical treatment for studying time dynamics. We make some simplifications of the full ME by using various useful identities.  These reduce ME  into a simple analytical form with decay rates corresponding to the various phonon-induced processes. Though we have not made any approximation, simplified ME scales down the numerical computation efforts and gives better insight into the physical process. By expanding all the commutators in Eq.(\ref{meq}) and rearranging using fermion operator identities, we get the simplified ME as
\begin{equation}
\begin{split}
\dot{\rho} &= \frac{1}{i\hbar}[H_s,\rho]+\sum\limits_{l}\Bigr(\frac{\gamma}{2}{\cal L}[\sigma^{-}_{l}]\rho +\frac{\gamma_d}{2}{\cal L}[\sigma^{+}_{l}\sigma^{-}_{l}]\rho\\ 
&+\frac{\Gamma^{\sigma^{+}}_{l}}{2}{\cal L}[\sigma^{+}_{l}]\rho
+\frac{\Gamma^{\sigma^{-}}_{l}}{2}{\cal L}[\sigma^{-}_{l}]\rho -\Gamma^{cd}_{l}(\sigma^{+}_{l}\rho\sigma^{+}_{l} + \sigma^{-}_{l}\rho\sigma^{-}_{l})\\
&- i\Gamma^{sd}_{l}(\sigma^{+}_{l}\rho\sigma^{+}_{l} - \sigma^{-}_{l}\rho\sigma^{-}_{l}) + i\Delta^{\sigma^{+}\sigma^{-}}_{l}[\sigma^{+}_{l}\sigma^{-}_{l},\rho]\\
&-[i\Gamma^{gu+}_{l}(\sigma^{+}_{l}\sigma^{-}_{l}\rho\sigma^{+}_{l} + \sigma^{-}_{l}\rho - \sigma^{+}_{l}\sigma^{-}_{l}\rho\sigma^{-}_{l})+H.c.]\\
&-[\Gamma^{gu-}_{l}(\sigma^{+}_{l}\sigma^{-}_{l}\rho\sigma^{+}_{l} - \sigma^{-}_{l}\rho + \sigma^{+}_{l}\sigma^{-}_{l}\rho\sigma^{-}_{l})+H.c.]\Bigr)\label{smeq}.
\end{split}
\end{equation}
The phonon-induced decay rates are given by 
\begin{widetext}
\begin{align}
\Gamma^{\sigma^{+} / \sigma^{-}}_{l} &= \frac{\Omega_R(z,t)^2}{2}\int_{0}^{\infty}\Bigg(\operatorname{Re}\bigg\{(\cosh(\phi(\tau))-1)f(z,t,\tau) + \sinh(\phi(\tau))\cos(\eta(z,t)\tau)\bigg\}\nonumber\\
&\mp \operatorname{Im}\left\{(e^{\phi(\tau)}-1)\frac{\Delta_{l}\sin(\eta(z,t)\tau)}{\eta(z,t)}\right\}\Bigg)\,d\tau, \label{phdrate1}\\
\Gamma^\mathrm{cd}_{l} &= \frac{1}{2}\int_{0}^{\infty} \operatorname{Re}\bigg\{\Omega_{S}(z,t)\sinh(\phi(\tau))\cos(\eta(z,t)\tau) - \Omega_{S}(z,t)(\cosh(\phi(\tau))-1) f(z,t,\tau)\nonumber\\
& + \Omega_{T}(z,t)(e^{-\phi(\tau)}-1) \frac{\Delta_{l}\sin(\eta(z,t)\tau)}{\eta(z,t)}\bigg\} d\tau, \label{gmcd}\\
\Gamma^\mathrm{sd}_{l} &= \frac{1}{2}\int_{0}^{\infty} \operatorname{Re}\bigg\{\Omega_{T}(z,t)\sinh(\phi(\tau))\cos(\eta(z,t)\tau) - \Omega_{T}(z,t)(\cosh(\phi(\tau))-1) f(z,t,\tau)\nonumber \\
& - \Omega_{S}(z,t)(e^{-\phi(\tau)} - 1) \frac{\Delta_{l}\sin(\eta(z,t)\tau)}{\eta(z,t)}\bigg\} d\tau,\\
\Delta^{\sigma^{+}\sigma^{-}}_{l} &= \frac{\Omega_R(z,t)^2}{2}\int_{0}^{\infty} \operatorname{Re}\bigg\{(e^{\phi(\tau)} - 1) \frac{\Delta_{l}\sin(\eta(z,t)\tau)}{\eta(z,t)}\bigg\} d\tau,\label{delpm}\\
\Gamma^{\mathrm{gu+}}_{l} &= \frac{\Omega_R(z,t)^2}{2}\int_{0}^{\infty}\bigg\{(\cosh(\phi(\tau))-1)\operatorname{Im}[\langle B\rangle\Omega]h(z,t,\tau)
+ \sinh(\phi(\tau))\frac{\operatorname{Re}[\langle B\rangle\Omega]\sin(\eta(z,t)\tau)}{\eta(z,t)}\bigg\}\,d\tau, \\
\Gamma^{\mathrm{gu-}}_{l} &= \frac{\Omega_R(z,t)^2}{2}\int_{0}^{\infty}\bigg\{(\cosh(\phi(\tau))-1)\operatorname{Re}[\langle B\rangle\Omega]h(z,t,\tau)
- \sinh(\phi(\tau))\frac{\operatorname{Im}[\langle B\rangle\Omega]\sin(\eta(z,t)\tau)}{\eta(z,t)}\bigg\}\,d\tau,\ \label{phdrate2}
\end{align}
\noindent where $f(z,t,\tau) = (\Delta_{l}^2\cos(\eta(z,t)\tau)+\Omega_R(z,t)^2)/\eta(z,t)^2$, $h(z,t,\tau) = \Delta_{l}(1 - \cos(\eta(z,t)\tau))/\eta^{2}(z,t)$ and $\eta(z,t) = \sqrt{\Omega_R(z,t)^2 + \Delta_{l}^2}$ with the polaron-shifted Rabi frequency, $\Omega_R(z,t) = \langle B\rangle\vert\Omega(z,t)\vert$, $ \Omega_{S}(z,t)=\operatorname{Re}[\langle B\rangle\Omega(z,t)]^{2}-\operatorname{Im}[\langle B\rangle\Omega(z,t)]^{2}$, $ \Omega_{T}(z,t)= 2\operatorname{Re}[\langle B\rangle\Omega(z,t)]\operatorname{Im}[\langle B\rangle\Omega(z,t)]$.
\end{widetext}

Next, we use Maxwell wave equation to describe the propagation dynamics of the electromagnetic field inside the QD medium
\begin{equation}
\bigg(\nabla^{2} - \frac{1}{c^{2}}\frac{\partial^{2}}{\partial t^{2}}\bigg)\vec{E}(z,t) = \mu_{0} \frac{\partial^{2}}{\partial t^{2}} \vec{P}(z,t)\label{wave_eq}
\end{equation}
where $\mu_{0}$ is the permeability of free space. The induced polarisation $\vec{P}(z,t)$ originates from the alignment of  the medium dipole in the presence of an applied field. Therefore it depends on the coherence term of the density matrix equation. For $l^{th}$ QD, the coherence term of the density matrix equation can be written as $\rho_{12}(\Delta_{l},z,t) = \langle 1 (\omega_{l})\vert_{l} \rho(z,t)\vert 2 (\omega_{l})\rangle_{l}$. The medium consists of a large number of QD with continuous frequency distribution centered at $\omega_{c}$. Therefore we can safely replace the summation with integration by redefining the discrete variable $\Delta_{l}$ to a continuous variable $\Delta$.
 The induced macroscopic polarisation can be written in terms of the density matrix element as 
\begin{equation}
\vec{P}(z,t) = N \int_{-\infty}^{\infty}\left(\vec{d}_{12}\rho_{12}(\Delta,z,t)e^{i(kz-\omega_{L} t)} + c.c.\right) g(\Delta)d\Delta,
\end{equation}
where $N$ is the QD volume number density. The inhomogeneous level broadening function in the frequency domain is defined by $g(\Delta)$.
In our calculation, the form of $g(\Delta)$ is
\begin{equation}
g(\Delta) = \frac{1}{\sigma\sqrt{2\pi}} e^{-\frac{(\Delta -\Delta_{c})^{2}}{2\sigma^{2}}},
\end{equation}
where the standard deviation is $\sigma$. The detuning between the applied field and the QD's central frequency is represented by $\Delta_{c}$.
By applying slowly varying envelope approximation, one can cast inhomogeneous second order partial differential Eq.(\ref{wave_eq})  to first order differential equation  as
\begin{equation}
\bigg(\frac{\partial}{\partial z} + \frac{1}{c}\frac{\partial}{\partial t} \bigg) \Omega(z,t) = i\hspace{1pt}\eta \int_{-\infty}^{\infty}\rho_{12}(\Delta,z,t)g(\Delta)d\Delta, \label{prop_eq}
\end{equation}
where the coupling constant $\eta$ is defined by 
\begin{equation}
\eta = - 3N\lambda^{2}\gamma/4\pi
\end{equation}
and $\lambda$ is the carrier wavelength of the QD transition. The self consistent solution of Eq.($\ref{smeq}$) and ($\ref{prop_eq}$) with proper initial conditions can display the spatiotemporal  evolution of the field inside the medium. Moreover the analytical solution of the coupled partial differential equation is known only for some special conditions, hence we adopted numerical integration of Eq.($\ref{smeq}$) and ($\ref{prop_eq}$) to depict the results. For numerical computation, a useful frame transformation $\tau = t - z/c$ and $\zeta = z$ is needed which removes the explicit time variable from Eq.($\ref{prop_eq}$), which now only depends on the one variable $\zeta$.

\section{NUMERICAL RESULT}
\label{sec:result}

\subsection{Phonon-induced scattering rates}

First we discuss various decay rates for the QD system with experimentally available parameter regions \cite{PARA1,PARA2}. 
The medium comprises InGaAs/GaAs QDs with volume density $N = 5\times 10^{20} \text{m}^{-3}$ and a length of 1\ mm. The central QD excitation energy is $\hbar\omega_{c}$ =1.3 eV with a Gaussian spectral distribution having FWHM of 23.5 meV. The QD is driven by the optical pulse at $\zeta$ = 0 with a hyperbolic secant profile
\begin{equation}
\vert\Omega(0,\tau)\vert = \Omega_{0}\hspace{1pt}\text{sech}\left(\frac{\tau-\tau_{c}}{\tau_{0}}\right)\label{rabi}
\end{equation}
where $\tau_{0}$, and $\tau_{c}$ defines the width, and center of the pulse, respectively. For numerical computation, the amplitude and width of the pulse are taken to be $\Omega_{0}$ = 0.2 meV and $\tau_{0}$ = 6.373 ps.
The phonon bath temperature  T = 4.2 K gives $\langle B \rangle = 0.95 $. Other parameters are $\alpha_{p} = 0.03\ \text{ps}^{2}$, $\omega_{b} = 1\ \text{meV}$. The system under consideration has a relaxation rate $\gamma =\gamma_{d} = 2\ \mu\text{eV}$(2 ns). In order to normalize all the system parameters to a dimensionless quantity we have chosen normalization frequency to be $\gamma_{n}$ = 1 rad/ps.

\begin{figure}[h]
   \includegraphics[scale=0.4]{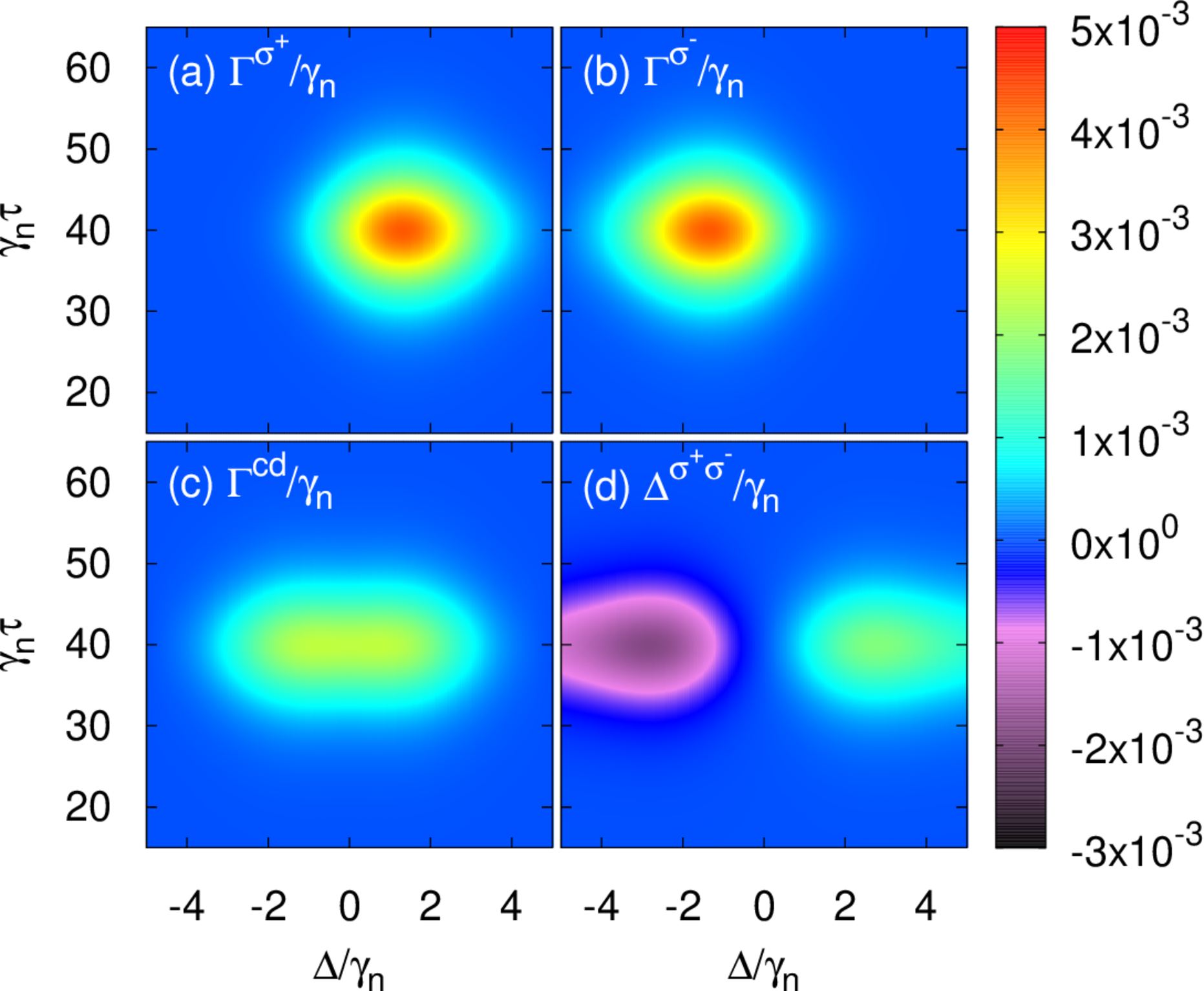} 
    \caption{The variation of phonon-induced scattering rates with detuning and time of a QD at $\zeta$ = 0 for the applied secant pulse in Eq.(\ref{rabi}). a) Phonon-induced pumping rate $\Gamma^{\sigma^{+}}$[Eq.(\ref{phdrate1})] b) Phonon-induced decay rate $\Gamma^{\sigma^{-}}$[Eq.(\ref{phdrate1})] c) Phonon induced dephasing $\Gamma^\mathrm{cd}$[Eq.(\ref{gmcd})] d) Phonon induced detuning $\Delta^{\sigma^{+}\sigma^{-}}$[Eq.(\ref{delpm})] for peak Rabi frequency $\Omega_{0}$ = 0.2 meV, pulse width $\tau_{0}$ = 6.373 ps and pulse center $\gamma_{n}\tau_{c} = 40$. The phonon bath temperature T = 4.2 K corresponds to $\langle B \rangle = 0.95 $ with spectral density function parameters $\alpha_{p} = 0.03\ \text{ps}^{2}$, $\omega_{b} = 1\ \text{meV}$.}
   \label{Fig.2}
\end{figure}

In Fig.(\ref{Fig.2}), the color bar represents the variation of various phonon-induced scattering rates as a function of  detuning and time, both at normalised units along the $x$- and $y$-axis respectively. In the QD system, various phonon processes are connected with exciton transitions. In the case of ground state to exciton transition, phonon absorption occurs while in the opposite process, phonon emission occurs.
Now we discuss the physical process associated with the phonon scattering rates $\Gamma^{\sigma^{+}}$ and $\Gamma^{\sigma^{-}}$. For positive detuning, the applied field frequency is larger than the QD transition frequency. Subsequently a  phonon generates with $\Delta$ frequency in order to make a resonant QD transition. These emitted phonons develop an incoherent excitation in the system referred by the $\Gamma^{\sigma^{+}}$. Oppositely for negative detuning, the applied field frequency is smaller than the QD transition frequency,  and a resonant QD transition is possible only  when some phonon of frequency $\Delta$ will be absorbed from the bath. With this mechanism, QD exciton to ground state decay enhances the radiation which is represented by the $\Gamma^{\sigma^{-}}$. This low-temperature asymmetry is clearly visible in Fig.$\ref{Fig.2}$(a) and $\ref{Fig.2}$(b). At higher temperatures, this  asymmetry gets destroyed, and  both rates overlap and are centered at $\Delta$ = 0. Fig.$\ref{Fig.2}$(c) shows the variation of $\Gamma^{cd}$ which is only present in the off-diagonal density matrix element and responsible for the additional dephasing in the system dynamics. The additional detuning $\Delta^{\sigma^{+}\sigma^{-}}$ from the simplified master equation plotted in Fig.$\ref{Fig.2}$(d), shows a very tiny value compared to the system detuning $\Delta$. We also notice that the sign of $\Delta^{\sigma^{+}\sigma^{-}}$ changes according to the system detuning $\Delta$. It is important to keep in mind that we display the variation along the $y$-axis around $\gamma_{n}\tau$ =40, which is the centre of the pulse with the secant profile.

\subsection{Pulse area theorem}
It is well know from Beer's law,  that a weak pulse gets absorbed inside the medium due to the presence of opacity at the resonance condition. However, McCall and Hahn showed that some specific envelope pulse shape remains intact for a long distance without absorption, even at resonance\cite{McHa1, McHa2}. Inspired of  this phenomena, we have taken into account of a time-varying pulse whose envelope shape is stated in the Eq.(\ref{rabi}). The area $\Theta(z)$ enclosed by its hyperbolic envelope shape is defined as
\begin{equation}
\Theta(z) = \int_{-\infty}^{+\infty}\Omega(z,t^{'})dt^{'}.
\end{equation}
By formally integrating Eq.($\ref{prop_eq}$) over time and detuning, one can find the spatial variation of the pulse area closely followed by the McCall and Hahn work.
The evolution of the pulse area $\Theta(z)$ during its propagation in a two-level absorbing QD medium is given by
\begin{equation}
\frac{d\Theta(z)}{dz} = -\frac{\alpha}{2} \sin\Theta(z)\label{paeq}
\end{equation}
where $\alpha$ is the optical extinction per unit length. The optical extinction depends on the various system parameters as $\alpha = 2\pi\eta g(0)$.  The solution of the Eq.(\ref{paeq}) is
\begin{equation}
 \tan \frac{\Theta(z)}{2} = \tan \frac{\Theta(0)}{2} e^{-\alpha z/2},\label{pasol}
\end{equation}
 where $\Theta(0)$ is the pulse area at $z$ =0. 
It is clear from the above expression that $\Theta(z) = 2n\pi$ is the stable solution, whereas $\Theta(z) = (2n+1)\pi$ is an unstable one.  The pulse area of the given envelope as stated in Eq.(\ref{rabi}) is $\Theta(0) = \pi\Omega_{0}\tau_{0}$. Thus, the envelope with amplitude $\Omega_{0} = 2/\tau_{0}$ gives  2$\pi$ area pulse. This envelope shape remains preserve for the long propagation distance even though it interacts resonantly  with the medium.\\
\begin{figure}[h]
   \includegraphics[scale=0.34]{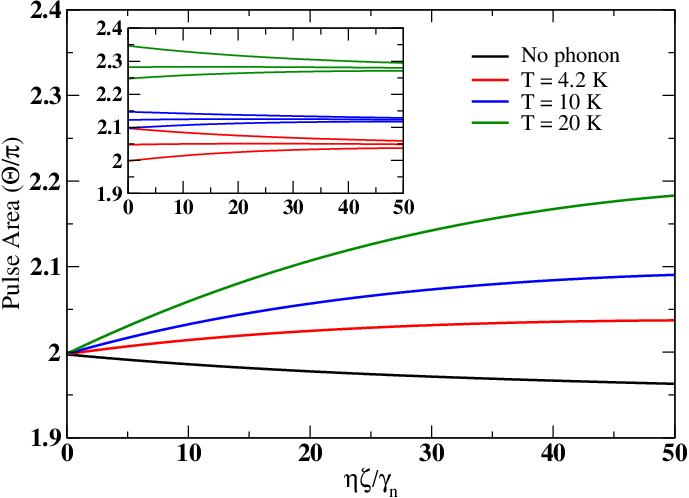}
    \caption{Evolution of the pulse area($\Theta$) as a function of propagation distance $\zeta$ started with $2\pi$ sech-type pulse for different temperatures. The applied pulse has a width of $\tau_{0}$ = 6.373 ps and centered at $\gamma_{n}\tau_{c} = 40$. The system under consideration without phonon bath(black) and with phonon bath maintaining temperature T = 4.2K(red), 10K(blue), 20K(green) with electron phonon coupling $\alpha_{p} = 0.03\ \text{ps}^{2}$ and cut off frequency $\omega_{b} = 1\ \text{meV}$.The central QD detuning $\Delta_{c}$ = 0 with spontaneous decay and the pure dephasing rate $\gamma =\gamma_{d} = 2\ \mu\text{eV}$(2 ns).The optical extinction per unit length $\alpha$ = 10 mm$^{-1}$. The inset figure shows the stability of the pulse area higher than $2\pi$ for different phonon bath temperatures.}
    \label{Fig.3}
\end{figure}
Fig.(\ref{Fig.3}) exhibits the variation of pulse area with the propagation distance inside the QD medium. It is evident from this figure that the propagation dynamics of 2$\pi$ area pulse through the medium of length $L$ has negligible loss in pulse area. In the absence of phonon(black line) interaction, the system behaves identical to the atomic system and hence follows $\Theta \approx 2\pi(1 - \tau_{0}/T_{2}^{'})$ reported earlier by McCall and Hahn \cite{McHa2}. The loss in pulse area comes from the finite lifetime $T_{2}^{'}$ of the QD which is inversely proportional to $\gamma_{d}$. Ideally, the pulse will retain initial pulse area for an arbitrary distance in absence of decay and decoherence. However, in presence of phonon contribution, we have noticed the pulse area gets enhanced by a small amount. The amount of raise in the  pulse area linearly depends on the bath temperature as indicated in Fig.($\ref{Fig.3}$).  This effect can be explained by carefully examing the definition of an effective Rabi frequency $\Omega_R(z,t) = \langle B\rangle\vert\Omega(z,t)\vert$ where $\langle B\rangle$ is dependent on the bath temperatures. The inset of Fig.(\ref{Fig.3}) illustrate the convergence of the pulse area shifted from the $2\pi$ value at different temperatures.\\
\begin{figure}[h]
   \includegraphics[scale=0.36]{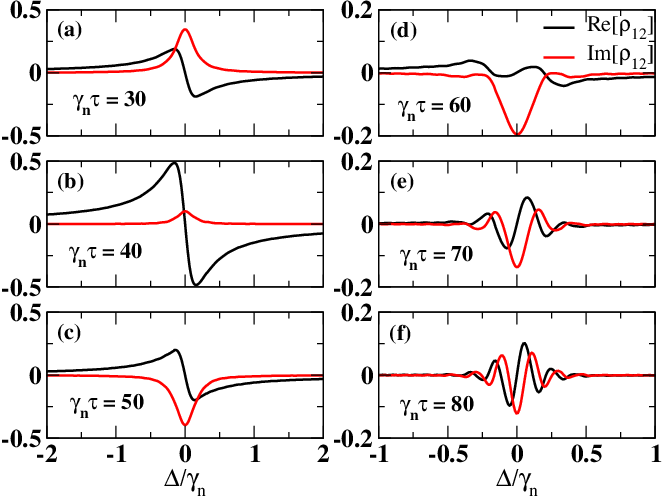}
    \caption{The real(black) and imaginary(red) part of the coherence $\rho_{12}$ of a single QD at different times for a $2\pi$ sech-type short pulse with pulse center at $\gamma_{n}\tau_{c}$ = 40 as a function of detuning. The pulse has a width $\tau_{0}$ = 6.373 ps. Corresponding phonon bath parameters are T = 4.2K, $\alpha_{p} = 0.03\ \text{ps}^{2}$, $\omega_{b} = 1\ \text{meV}$. Considered QD relaxation rates are $\gamma =\gamma_{d} = 2\ \mu\text{eV}$(2 ns). }
    \label{Fig.4}
\end{figure}
To explain the behavior of Fig.(\ref{Fig.3}), we study the absorption and dispersion properties of the medium as a function of detuning at various time intervals of the pulse. Fig.(\ref{Fig.4}) delineates the physical process behind the dispersion and absorption. We assume all the population in the ground state, before the leading edge of the pulse reaches the medium. The peak of incident pulse enters inside the medium at $\gamma_{n}\tau_{c}$ = 40. It is clear from Fig.\ref{Fig.4}(a) that most of the leading edge pulse energy gets absorbed by the ground state population and the population goes to the excited state. Hence the medium shows maximum absorption at $\gamma_{n}\tau$ = 30, hence elucidating the absorption phenomenon at resonance. Simultaneously, the nature of the dispersion curve is anomalous as previously reported \cite{BOYD}. The anomalous dispersion accompanied fast velocity is completely prohibited due to huge absorption at the resonance condition. The medium becomes saturated as the centre of the pulse enters the medium; consequently, the medium turns less absorbent to the pulse.  Nonetheless, a tiny absorption peak still exists at the resonance condition due to the presence of various decay processes of the medium as indicated by Fig.\ref{Fig.4}(b).  Therefore, the excited state gets populated during the passage of the leading edge pulse. This population can leave the excited state and return to the ground by stimulated emission in the presence of the trailing edge of the pulse. As a results, a gain can be experienced by the incident pulse at $\gamma_{n}\tau$ = 50 as revealed in Fig.\ref{Fig.4}(c). From these three panels, we can conclude that the leading edge of the pulse gets absorbed by the medium, while the tailing edge of the pulse experiences gain. Towards the trailing end of the pulse, the dispersive nature of the medium changes from anomalous to normal, as shown in Fig.\ref{Fig.4}(d). The positive slope of the dispersion curve lead to slow group velocity that started at $\gamma_{n}\tau$ = 60 shown in Fig.\ref{Fig.4}(d). Fig.\ref{Fig.4}(d)  to Fig.\ref{Fig.4}(f) indicate that the optical pulse regeneration process is completed due to the medium-assisted gain; hence, the pulse shape remains preserved. This is the explanation of the underpinning mechanism behind SIT.
\begin{figure}[h]
   \includegraphics[scale=0.33]{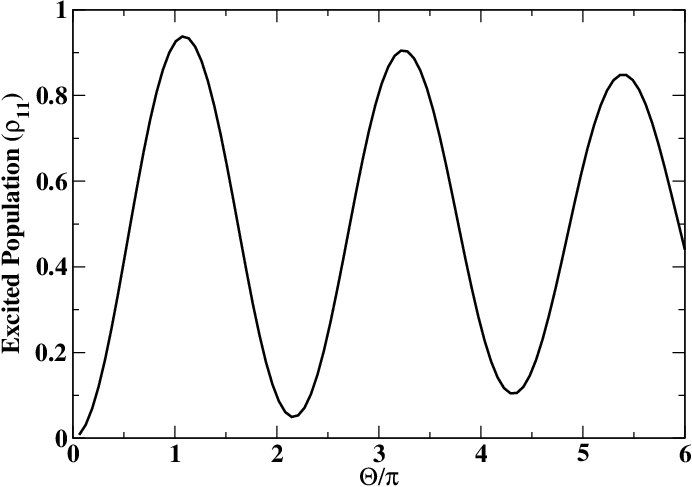}
    \caption{The variation of excited state population with input pulse area at resonance condition $\Delta_{c}$ = 0. The system and bath parameters are $\tau_{0}$ = 6.373 ps, $\gamma_{n}\tau_{c}$ = 40, T = 4.2K, $\alpha_{p} = 0.03\ \text{ps}^{2}$, $\omega_{b} = 1\ \text{meV}$, $\gamma =\gamma_{d} = 2\ \mu\text{eV}$(2 ns).}
    \label{Fig.5}
\end{figure}
The claim of the above physical mechanism can be supported by studying population dynamics at the excited state. For this purpose, we have plotted the excited state population as a function of the pulse area in Fig.(\ref{Fig.5}). A noticeable population redistribution among the levels is feasible within few widths of incident pulse wherein intensity is appreciable. As soon as the pulse intensity diminishes at the trailing end, spontaneous emission takes care of depletion of the excited state population. This leads to vanishing population at the excited state after a sufficiently long time from the pulse centre. As a consequence, it is crucial to decide the observation time of the QD population. Hence, we display the exciton population at just the end of the pulse $\gamma_{n}\tau = 60$, to capture the outcome of the pulse. It is clear from Fig.(\ref{Fig.5}) that the excited state population shows a decaying Rabi oscillation kind of behaviour. It is also confirmed that the population never fully transferred to the excited state or fully returned to the ground state for any pulse area, indicating to non-constant phonon induced decay and gain process involved in the system. The decaying features of the local population maximum can be justified by the examining the photon and phonon induced decay rates. The various phonon decay  rates are given in Eq.(\ref{phdrate1})-(\ref{phdrate2}) where increasing incident pulse amplitude $\Omega(z,t)$ results in the enhancement of these decay rates.  This field amplitude dependent phonon decay together with constant photon decay can explain the gradual decay of the population local maxima. On the contrary, the dip of local minima increases due to the presence of phonon induced gain processes $\Gamma^{\sigma+}$ as suggested in Eq.(\ref{phdrate1}).  The local maximum and minimum of the exciton population are located respectively near odd and even integer multiples of $\pi$ pulse area. The maxima signifies the pulse absorption by the medium, resulting in population inversion. Similarly, minima manifests the transparency of the medium. Thus, the leading edge of the pulse excite the population whereas the tailing edge assists in stimulated emission leaving the population in the ground state of the medium. It is evident that only even integer multiples of $\pi$ pulse can propagate through the medium without absorption that is consistent with the pulse area theorem. That the local maxima and minima of exciton population never match exactly with the integer value can be figured out later by investigating pulse propagation dynamics. Previously, we found the stable pulse area is higher than $2\pi$ as shown in Fig.($\ref{Fig.3}$) which also agrees with the above observation.  Therefore, the analysis of coherence and population ensures us that SIT phenomena can be accomplished in the QD medium.
\begin{figure}[h]
   \includegraphics[scale=0.33]{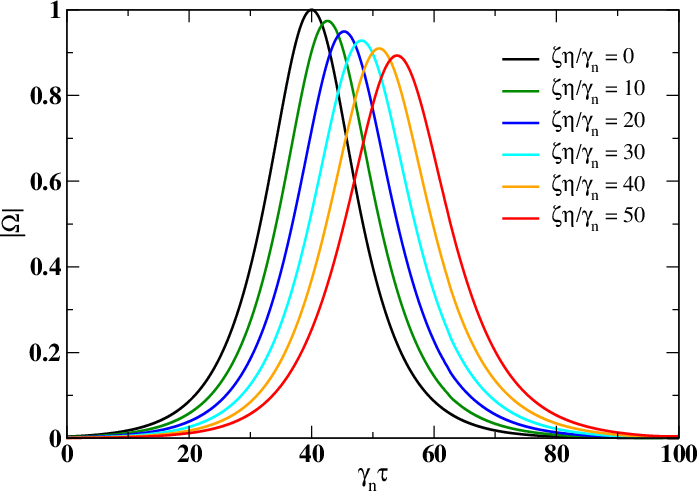}
    \caption{The Rabi frequency normalized with the input peak value is  plotted against retarded time at different propagation distances inside the medium at resonance condition $\Delta_{c}$ = 0. The input pulse has the following parameters $\Theta(0) = 2\pi$, $\tau_{0}$ = 6.373 ps, $\gamma_{n}\tau_{c}$ = 40. The chosen QD inhomogeneous level broadening in normalized units $\sigma/\gamma_{n} = 15$. Other parameters are T = 4.2K, $\alpha_{p} = 0.03\ \text{ps}^{2}$, $\omega_{b} = 1\ \text{meV}$, $\gamma =\gamma_{d} = 2\ \mu\text{eV}$(2 ns).}
    \label{Fig.6}
 \end{figure}   
\subsection{Self Induced Transparency}
A homogenous QD medium with length $1$\ mm is taken into account for studying spatio-temporal evolution of hyperbolic secant optical pulse. To achieve a stable pulse propagation, we have chosen the initial pulse area to be 2$\pi$. Fig.$\ref{Fig.6}$ confirms the area theorem by showing a stable optical pulse propagation for a longer distance. However, the pulse shape at larger distances has noticed some distortion and absorption.
Fig.$\ref{Fig.6}$ also indicates that the pulse's peak value gradually decreases by increasing the propagation distance. This suggests a finite absorption in the QD medium that prohibited complete transparency in the system.  In particular, the statement agrees well  with the small absorption peak at resonance in the absorption profile shown in Fig.$\ref{Fig.4}$b.
\begin{figure}[h]
   \includegraphics[scale=0.33]{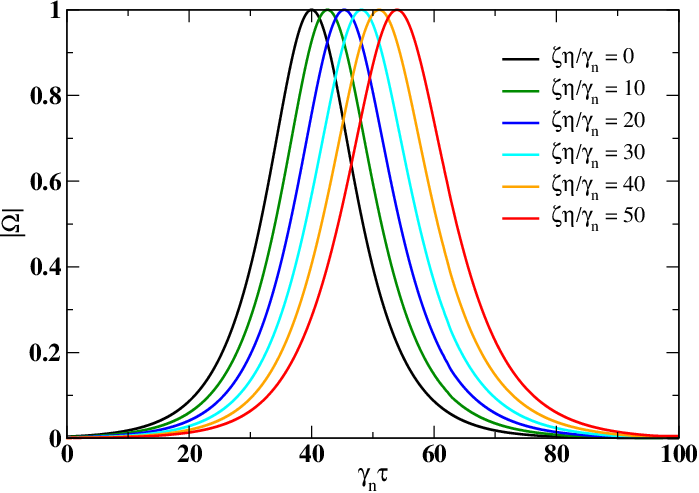}
    \caption{The Rabi frequency normalized with the individual peak value is  plotted against retarded time at different propagation distances inside the medium at resonance condition $\Delta_{c}$ = 0. All the other parameters are the same as Fig.(\ref{Fig.6}).}
    \label{Fig.7}
\end{figure}
Figure \ref{Fig.7} displays the individually normalized pulse for different propagation distances. Inspection says that the input pulse experiences delay and a little broadening during the propagation through the medium. The sole reason behind the pulse broadening is the dispersive nature of the system. In the frequency domain, a temporal pulse can be treated as a linear superposition of many travelling plane waves with different frequencies. These individual frequency waves gather different phases and move with varying velocities during the pulse propagation in a dispersive medium. Therefore the pulse gets broader as the leading part(low frequency) moves faster, and the tailing end(high frequency) goes slower. In the QD system, the pure dephasing rate is also responsible for this broadening as it destroys the coherence.
From Fig.$\ref{Fig.7}$, a distinct peak shift is observed while optical pulse propagating through the medium. This peak shift arises because of normal dispersive medium that induced slow group velocity of the optical pulse inside the medium.
We adopt the analytical expression of time delay in the ideal case by considering $\sigma \gg 1/\tau_{0}$ reported earlier\cite{RAHMAN}. The  analytical expression for time delay found to be $\gamma_{n}\tau_{d} = \alpha L \gamma_{n}\tau_{0}/4$. Here the absorption coefficient $\alpha$ is approximately 10 mm$^{-1}$ calculated from the chosen parameters. Therefore the calculated  analytical time delay $\gamma_{n}\tau_{d}\approx$ 15 shows excellent agreement with the numerical result.
The inhomogeneous level broadening $\sigma$ plays an important role in pulse propagation dynamics. In our calculation, we are in the regime where the pulse width is greater than the inhomogeneous broadening time $\sigma \tau_{0}\gg 1$. Therefore, the higher spread of the QD parameter $\sigma$  leads to fewer QD resonantly interacting with the propagating pulse. This results in a negligibly small change in pulse shape. Alternatively, the effective QD density becomes less, indicating the lower value of the optical extinction parameter $\alpha$. Henceforth a lower time delay is expected in the final output pulse due to its presence in the righthand term of Eq.(\ref{prop_eq}). In Fig.($\ref{Fig.8}$), the final output pulse shape variation is presented for the three different QD spreads. The pulse delay decreases with an increasing QD broadening $\sigma$. On the other hand, the pulse peak value decreases with the lower QD spreads. This observation matches our previous prediction that higher $\sigma$ produce a lower pulse delay in the medium. Also, more resonant QD absorb more energy from the pulse, resulting in a lesser peak value in the final pulse shape. Hence spread of the QD is also a determining factor for the shape and delay of the output pulse.
\begin{figure}[h]
   \includegraphics[scale=0.33]{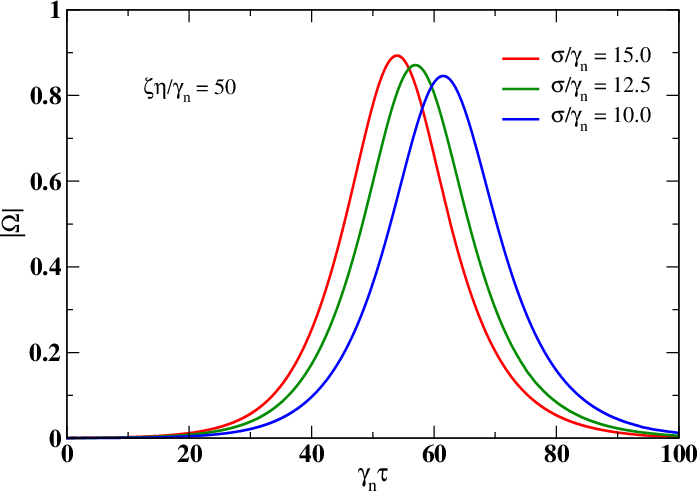}
    \caption{The normalized  Rabi frequency displayed with retarded time after passing the medium for three different QD broadening $\sigma$. All the other parameters are the same as Fig.(\ref{Fig.6}).}
    \label{Fig.8}
\end{figure}
Recalling the pulse area theorem again, we observe that the pulse area is almost constant throughout the propagation near $2\pi$. The result is consistent because as the pulse amplitude decreases, the pulse width increases, maintaining the constant area under the curve. Therefore an absorbing QD medium can exhibit the SIT phenomena at low temperatures.
 
\subsection{Phonon bath parameter dependence on SIT }
In the simplified master equation ($\ref{smeq}$), various phonon-induced scattering rates depend on both the system and bath parameters. Hence it is crucial to study the effect of phonon bath on the SIT dynamics. The phonon contribution comes to the picture in two ways; one from the reduced Rabi frequency, which depends on the $\langle B \rangle$ and the other is the phonon-induced scattering rates connected with the phonon spectral density function.
\begin{figure}[h]
   \includegraphics[scale=0.33]{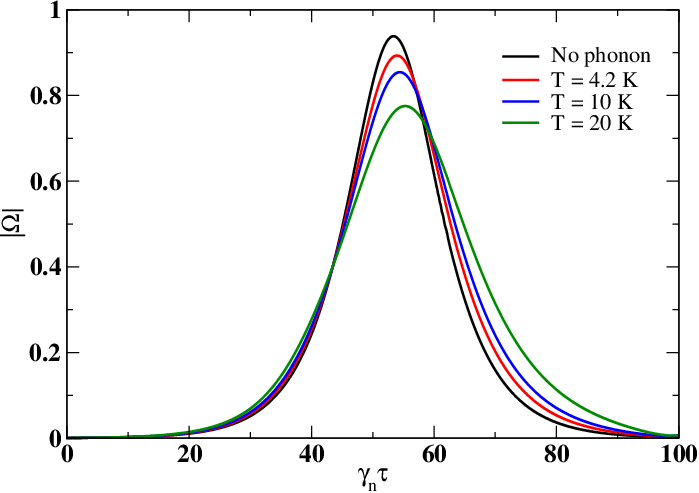}
    \caption{The plot of Rabi frequency envelope with time at a propagation distance $\zeta\eta/\gamma_{n}$ = 50 for different phonon bath temperatures at resonance condition $\Delta_{c}$ = 0. The common parameters are  $\Theta(0) = 2\pi$, $\tau_{0}$ = 6.373 ps, $\gamma_{n}\tau_{c}$ = 40, $\alpha_{p} = 0.03\ \text{ps}^{2}$, $\omega_{b} = 1\ \text{meV}$, $\gamma =\gamma_{d} = 2\ \mu\text{eV}$(2 ns). The figure display four different configurations, system without a phonon bath (black) and with a phonon bath at a temperature T = 4.2K(red), 10K(blue), 20K(green).}
    \label{Fig.9}
\end{figure}
Therefore increasing phonon bath temperatures reduces the value of $\langle B \rangle$ and $\hbar\omega/2K_{b}T$ present in the expression of $\phi(\tau)$ given in the Eq.($\ref{phi}$). Consequently, effective coupling between QD and applied field gets reduced, but the phonon-induced decay rates get enhanced. From Fig.($\ref{Fig.9}$), we notice that the final pulse shape experiences more deformation for higher temperatures. The peak of the output pulse is also very much reduced for the higher temperature T =20 K. Therefore, the bath temperature should be minimised to see the SIT in the QD medium.
\begin{figure}[h]
   \includegraphics[scale=0.33]{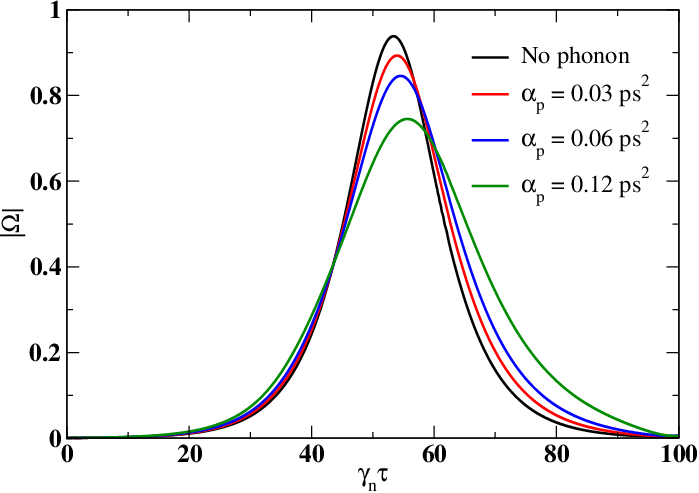}
    \caption{The Rabi frequency envelope with time at a propagation distance $\zeta\eta/\gamma_{n}$ = 50 for different electron-phonon coupling strength $\alpha_{p}$ at resonance condition $\Delta_{c}$ = 0. All the parameters are same as Fig.(\ref{Fig.9}) except T = 4.2K and various electron-phonon coupling $\alpha_{p} = 0.03\ \text{ps}^{2}$(red), 0.06\ $\text{ps}^{2}$(blue), 0.12\ $\text{ps}^{2}$(green).}
    \label{Fig.10}
\end{figure}
Another controlling factor of the SIT is the interaction strength between the QD and the phonon bath. So the increment of system-bath coupling leads to the reduction of the coherence in the system. This statement is understandable by looking at the phonon correlation function shown in Eq.($\ref{phi}$). Thus the final pulse shape for the equal propagation distances is significantly modified by the electron-phonon coupling constant, as shown in Fig.($\ref{Fig.10}$). Therefore we also have to ensure that the QD bath interacts weakly to get SIT phenomena in the QD medium.
\subsection{Higher pulse area and pulse breakup}
\begin{figure}[h]
   \includegraphics[scale=0.75]{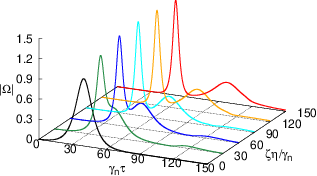}
    \caption{The propagation dynamics of a 4$\pi$ area pulse in an absorbing QD medium as a function of both space and time at resonance condition $\Delta_{c} = 0$. All other parameters are same as Fig.(\ref{Fig.6}).}
    \label{Fig.11}
\end{figure}

Finally, we discuss the behaviour of a pulse propagating through the absorbing QD medium with a higher pulse area than $2\pi$. Therefore we consider the next stable pulse area solution 4$\pi$ for further investigation. The numerical result of the pulse propagation in both space and time is shown in Fig.($\ref{Fig.11}$). Unlike the 2$\pi$ pulse case, here, the initial pulse breaks into two pulses as it travels through the medium. This phenomenon is also well explained by the pulse area theorem where $2n\pi$ pulse is split into $n$ number of 2$\pi$ pulses. Surprisingly, the initial pulse breakup into two pulses is not identical in shape. One pulse gets sharper, and the other gets broader in the time domain and adjusts the peak value such that the area under the curve is 2$\pi$. The broader pulse component shows a prominent time delay, whereas the sharper pulse component propagates with a tiny time delay. As a result, total pulse area is constant throughout the propagation distance near 4$\pi$.
\section{CONCLUSIONS}
\label{sec:conclud}
We have investigated the SIT phenomena in an inhomogeneously broadened semiconductor QD medium. In our model, we have included the effect of phonon in the total Hamiltonian to describe the modified optical properties of QD in the presence of a thermal environment. We then adopted the polaron ME formalism to analytically derive the simplified ME with various phonon-induced decay rates. These phonon-induced scattering rates are plotted against detuning and time, which verify the presence of low-temperature asymmetry of phonon-induced pumping and decay in our system. We solve numerically the density matrix equation and Maxwell equation selfconsistently with suitable parameters. We observe that stable pulse propagation is possible in the QD medium with pulse area slightly higher  than 2$\pi$, depending on the phonon bath temperature. The physical mechanism of the SIT is clearly understood by analyzing the absorption and dispersion of the medium. The leading edge of the pulse gets absorbed by the medium, whereas the tailing edge of the pulse experience gain, hence the pulse shape remains intact and propagate through  medium with short length. However, for longer propagation distances, we find that even though the pulse propagation through the medium is possible, the propagating pulse gets absorbed and broadened. The final pulse shape is preserved on exiting the medium. Increasing the phonon bath temperature and coupling produce more deformation in the final pulse shape, as it destroys the coherence in the system. Finally, we explore the propagation of a 4$\pi$ pulse in the QD medium, which shows prominent pulse breakup phenomena reported earlier in the literature.
Therefore our investigation ensures that a short pulse can propagate through the considered QD medium with a tiny change in shape. Hence, this work may have potential applications in quantum communication, quantum information, and mode-locking.

\bibliography{Paper}

\end{document}